\documentclass[aps,prd,showpacs,nofootinbib,floats,floatfix,preprintnumbers,groupedaddress,twocolumn]{revtex4}

\usepackage{bm}
\usepackage{latexsym}
\usepackage{dcolumn}
\usepackage{amsmath,amsfonts,amssymb}
\usepackage{graphicx,epsfig}

\usepackage{fancyhdr}

\def\beq{\begin{equation}}
\def\eeq{\end{equation}}
\def\bea{\begin{eqnarray}}
\def\eea{\end{eqnarray}}
\def\benu{\begin{enumerate}}
\def\eenu{\end{enumerate}}

\def\lp{L_{_{\rm P}}}

\def\lgb{\alpha_{_{\rm {GB}}}}

\reversemarginpar

\newcommand{\LL}{Lanczos-Lovelock }
\newcommand{\Cal}[1]{\ensuremath{\mathcal{#1}}}
\newcommand{\dV}{\ensuremath{\partial\mathcal{V}}}
\newcommand{\D}{\ensuremath{\nabla}}

\newcommand{\ph}[1]{\phantom{#1}}

\def\eq#1{{Eq.~(\ref{#1})}}

\begin{document}
\title{Is gravitational entropy quantized ?}
\author{Dawood Kothawala}
\email{dawood@iucaa.ernet.in}
\author{T.~ Padmanabhan}
\email{paddy@iucaa.ernet.in}
\author{Sudipta Sarkar}
\email{sudipta@iucaa.ernet.in}
\affiliation{IUCAA,
Post Bag 4, Ganeshkhind, Pune - 411 007, India}

\date{\today}
\begin{abstract}
In Einstein's gravity, the entropy of horizons is proportional to their area. Several arguments given in the literature suggest that, in this context, both area and entropy should be quantized with an equally spaced spectrum for large quantum numbers. But in more general theories (like, for e.g, in the black hole solutions of Gauss-Bonnet or \LL gravity)  the horizon entropy is \textit{not} proportional to area and the question arises as to which of the two (if at all) will have this property. We give a general argument that in all \LL theories of gravity, it is the \textit{entropy} that has equally spaced spectrum. In the case of Gauss-Bonnet gravity, we use the asymptotic form of quasi normal mode frequencies to explicitly demonstrate this result. Hence, the concept of a quantum of area  in Einstein Hilbert (EH) gravity needs to be replaced by a concept of \textit{quantum of entropy} in a more general context.

\end{abstract}  \maketitle



It was conjectured by Bekenstein \cite{Bekenstein1} long back that, in a quantum theory, the black hole {\it area} would be represented by a quantum operator with a discrete spectrum of eigenvalues.
 Bekenstein showed that the area of a classical black hole behaves like an adiabatic invariant, and so, according to Ehrenfest's theorem, the corresponding quantum operator must have a discrete spectrum. It was also known that, when a quantum particle is captured by a (non extremal) black hole its area increases by a minimum non-zero value \cite{chris,Bekenstein1,comment}  which is independent of the black hole parameters. This argument also suggests an equidistant spacing of area levels, with a well-defined notion of a {\it quantum of area}. The fundamental constants $G,c$ and $\hbar$ combine to give a quantity with the dimensions of area $A_P=(G \hbar / c^3)=10^{-66}$ cm$^2$, which is quite suggestive \cite{zeropoint} and sets the scale in area quantization.

In Einstein's gravity, entropy of the horizon is proportional to its area. Hence one could equivalently claim that it is the gravitational entropy which has an equidistant spectrum with a well-defined notion of \textit{quantum of entropy}.
But, when one considers the natural generalization of Einstein gravity by including higher derivative correction terms to the original Einstein-Hilbert action, no such trivial relationship remains valid between horizon area and associated entropy. One such higher derivative theory which has attracted a fair amount of attention is Lanczos-Lovelock (LL) gravity \cite{lovelock}, of which the lowest order correction appears as a Gauss-Bonnet (GB) term in $D(> 4)$ dimensions. These lagrangians have the unique feature that the field equations obtained from them are quasi linear, as a result of which, the initial value problem remains well defined. More importantly, several features related to horizon thermodynamics, which were first discovered in the context of Einstein's theory \cite{gravtherm}, continues to be valid in LL gravity models \cite{aseem-sudipta,ayan}.

Black hole solutions in the LL gravity are well studied in the literature. For these spacetimes, the notion of entropy can be defined using Wald's formalism \cite{noether}, where entropy is associated with the Noether charge of the diffeomorphism invariance symmetry of the theory. The entropy calculated from this approach turns out to be no longer proportional to horizon area. The question then arises as to whether it is the quantum of area or quantum of entropy (if at all either) which arises in a natural manner  in these models.
We attempt to answer this question in this paper.

We will first provide a very general argument which suggests that it is the entropy which is quantized with equidistant spectrum in the case of LL gravity and then provide an explicit proof for the result in the context of GB gravity.

In any geometrical description of gravity that obeys the principle of equivalence and is based on a nontrivial metric, the propagation of light rays will be affected by gravity. This, in turn, leads to regions of spacetime which are causally inaccessible to classes of observers. (These two features are reasonably independent of the precise field equations which determine the metric.). The fact that any observer has a right to formulate physical theories in a given coordinate system entirely in terms of the variables that an observer using that coordinate system can access, imposes strong constraints on the nature of the action functional  $A_{grav}$ which can be used to describe gravity \cite{paddypatel}.  Suppose we divide the space-time manifold into two regions separated
by a null hypersurface $\mathcal{H}$ and choose a coordinate system such that $\mathcal{H}$ acts as a horizon for the observer on one side (say side 1).
The effective theory for the observer on side 1 (with the degrees of freedom formally denoted by $g_1$) is obtained by integrating out the variables on the inaccessible side (side 2):
In the semiclassical limit, saddle-point integration  leads to the exponential of the classical action $A_{\rm grav}({\rm class})$ evaluated on-shell.
The effective theory on side 1 is thus described by the action
$A_{\rm eff}^{^{\rm WKB}}(g_1)$, with
\begin{equation}
\exp [iA_{\rm eff}^{^{\rm WKB}}(g_1)] \simeq
\exp [i(A_{\rm grav}(g_1) + A_{\rm grav}({\rm class}))].
\end{equation}
Since the effects of the unobserved degrees of freedom $g_2$ can only be encoded in the geometry of the (shared) boundary between regions 1 and 2, we get the constraint that the on-shell value of the action $A_{\rm grav}(g_2^{\rm class})$
must be expressible in terms of the boundary geometry which could be expressible in terms of $g_1$ itself.
That is, $A_{\rm grav}(g_2^{\rm class})=A_ {sur}(g_1)$, and
\begin{equation}
\exp [iA_{\rm eff}^{^{\rm WKB}}(g_1)] =
\exp [i(A_{\rm grav}(g_1) + A_ {sur}(g_1))].
\label{thecondition}
\end{equation}
This is a  non-trivial requirement on any geometrical theory of gravity.
Further since the boundary term depends on the choice of the
coordinate system (or foliation), in which $\mathcal{H}$ acts as a one-way membrane,
$A_ {sur}(g_1)$ will in general depend on the coordinate choice for the observer.

Classically, with the boundary variables held fixed, the equations of
motion remain unaffected by the existence of a (total divergence) boundary term;
hence the fact that the boundary term is not generally covariant is
unimportant for \textit{classical} theory.
This is, of course, not true in semiclassical/quantum theory.
But since the quantum theory is governed by $\exp[iA_{\rm eff}]$
rather than  by $A_{\rm eff}$, the boundary term will have no effect in the quantum theory,
if the quantum processes keep $\exp[iA_ {sur}]$ single-valued.
This is equivalent to demanding that the boundary term satisfies the quantization condition $ A_ {sur}=2\pi n$. (More precisely, the change in the surface term $\Delta A_ {sur}=2\pi$; this is irrelevant for our purpose when we work in the semiclassical limit of large $n$).

It is now worth noting that the lagrangian in \textit{all the LL models} (of which Einstein-Hilbert action is just a special case) can be expressed \cite{ayan} as a sum of a bulk and total divergence terms, $L=L_{bulk}+L_{sur}$ with  $L_{sur}$ integrating to give a surface term in the action.
 There is a peculiar `holographic' relationship between $L_{bulk}$ and $L_{sur}$ in all these models with the same information being coded in both the bulk and surface terms [see eq.(41) of ref.\cite{ayan}].
The on-shell value of the surface term \textit{in all these action functionals} is proportional to the Wald entropy of the horizon \cite{ayan,aseementropy}. We can now see how a condition like $ A_ {sur}=2\pi n$ can lead to quantization of Wald entropy.

In the case of Einstein-Hilbert action, $A_ {sur}$ is well defined and is given by the standard Gibbons-Hawking-York term. As pointed out in ref.\cite{paddypatel} the surface term will give the entropy --- equal to one quarter of horizon area --- and both will have an equally-spaced spectrum. When we proceed to general LL gravity, the correct (surface) counterterm which should be added to the higher derivative action is unknown and the action principle, using metric as dynamical variable, is actually ill-defined. There is, however, an alternative approach we can follow to obtain meaningful results for the LL gravity.

So far we did not have to specify the exact nature of the degrees of freedom $g_1, g_2$ in the above discussion.
Interestingly enough these arguments go through unhindered, when one
 formulates gravity as an emergent phenomenon without treating the metric as dynamical variables in the theory \cite{emergentpaddy}. In this approach one proceeds along the following lines:

 Around any event in spacetime one can introduce a local inertial frame and --- by boosting with a uniform acceleration --- a local Rindler frame. The effective long range variables in emergent gravity approach are the normals $n_a$ to the null surfaces which act as local Rindler horizons. (One can think of $n_a$ as the `fluid velocity' of a virtual null fluid in the spacetime.). The total action is now \cite{emergentpaddy,aseementropy} taken to be $A_{\rm tot} = A_{\rm grav} + A_{\rm matt}$ where
 \begin{equation}
A_{grav}= - 4\int_\Cal{V}{d^Dx\sqrt{-g}}
    P_{ab}^{\ph{a}\ph{b}cd} \D_cn^a\D_dn^b
    \label{Sgrav}
\end{equation}
 is determined by a fourth rank tensor $P_{abcd}$ which can be expressed as a derivative of the  LL  lagrangian and
 \begin{equation}
A_{\rm matt}=\int_\Cal{V}{d^Dx\sqrt{-g}}
      T_{ab}n^an^b
      \label{Smatt}
\end{equation}
Maximizing $A_{tot}$ with respect to all $n_a$ leads to the field equations of the LL theory.
(All these aspects are described in detail in ref.\cite{aseementropy} and hence not repeated here.)
 The key result we need here is that the \textit{on-shell} value of the total action is given by
\begin{equation}
A^{tot}|_{\rm on-shell}=4\int_{\dV}{d^{D-1}\Sigma_a\left(P^{abcd} n_c\D_b n_d\right)}
\label{on-shell-2}
\end{equation}
which can be shown to be identically equal to the Wald entropy of the horizon in the  LL  theory \cite{ayan,aseementropy}.
(In the case of lowest order  LL  theory --- which is just Einstein gravity --- the expression for $A|_{\rm on-shell}$ will be one quarter of the transverse area of the horizon.)
 The emergent gravity approach is strongly motivated by thermodynamic considerations and --- \textit{classically} --- the maximization of the action can be thought of as maximization of the entropy. In this context, \eq{on-shell-2} will also give the entropy of the local Rindler horizon for each Rindler observer, which can be interpreted as due to integrating out the inaccessible degrees of freedom behind the local Rindler horizon.
In the \textit{semiclassical} limit, the
 on-shell value of the action will be related to the phase of the semiclassical wave function
$
\Psi \propto \exp \left(iA|_{\rm on-shell}\right)
$
This expression, of course, should be generally covariant but as it stands it explicitly depends on the Rindler observer chosen to define the horizon. Hence we can ensure observer independence of semiclassical gravity only if we assume
\begin{equation}
A_{\mathrm{Wald}}=A|_{\rm on-shell}=2\pi n
\end{equation}

Note that we are again obtaining the quantization condition from the phase of the semiclassical wavefunction which is completely in accord with previous approaches to this problem. The holographic relation between surface and bulk terms underscores how the surface term captures the dynamical information contained in the bulk.

While this gives a general result that in  LL  theories it is the entropy of the horizon that is quantized,  it would be nice if the result could be reinforced by an explicit calculation within the standard context.
Fortunately, this can be done for GB theory using the arguments suggested by Hod \cite{hod} based on quasinormal modes of black hole oscillations.

Hod started from Bekenstein's arguments regarding quantum area spectrum of a non extremal Kerr-Newman black hole, and showed that the spacing of area eigenvalues can be fixed by associating the classical limit of the quasinormal mode frequencies, $\omega_c$, with the large $n$ limit of the quantum area spectrum, in the spirit of Bohr's correspondence principle ($n$ being the quantum number). Specifically, for a Schwarzschild black hole of mass $M$ in $(3+1)$ dimensions, the absorption of a quantum of energy $\omega_c$, (in units with $\hbar=1$) would lead to change in the black hole area  eigenvalues as,
$
\Delta{\mathcal A}\equiv {\mathcal A}_{n+1} - {\mathcal A}_n = (\partial {\mathcal
A}/\partial M)   \omega_c
$
and for entropy $\Delta S=(\partial S/\partial M)   \omega_c$. In the case of a Schwarzschild black hole
the level spacing of both area and entropy eigenvalues were indeed found to be equidistant, allowing one to associate the notion of a minimum unit, {\it a quantum}, of area \textit{and} entropy.

   We will use these ideas in the context of $5D$ GB black holes, using the numerically known form of the quasinormal mode frequencies. We show that, the form of the highly damped quasi normal modes of these black holes suggest that it is the entropy which has a equally spaced spectrum.
The Gauss-Bonnet (GB) Lagrangian $L$ in $D$ dimensions is given by \cite{lovelockblack},
\begin{eqnarray*}
(16 \pi G)L= \left[ R  + \lgb (R^2 - 4 R_{ab}R^{ab}   +  R_{abcd}R^{abcd}) \right]
\end{eqnarray*}
Static, spherically symmetric black hole solutions in this theory is of the form,
\begin{eqnarray*}
{\mathrm d}s^2 &=& - f(r) {\mathrm d}t^2 + f(r)^{-1}{\mathrm d}r^2
+
r^2 {\mathrm d}\Omega_{D-2}
\end{eqnarray*}
where,
\begin{eqnarray*}
f(r) &=& 1 + \frac{r^2}{ 2 {\alpha} } \left[ 1 - \left( 1 +
\frac{4 ~ \alpha ~ \varpi}{r^{D-1}} \right)^{1/2}
\right]
\end{eqnarray*}
Here, $\alpha=(D-3)(D-4) \lgb$ and $\varpi$ is related to the ADM mass $M$ by the relationship,
\begin{eqnarray*}
\varpi &=& \frac{ 16 \pi G }{ (D-2) \Sigma_{D-2} } ~ M
\end{eqnarray*}
where $\Sigma_{D-2}$ is the volume of unit $(D-2)$ sphere. The Hawking temperature $T$, and entropy $S$ for this spacetime are,
\begin{eqnarray*}
T &=& \frac{D-3}{4 \pi r_{+}} \Biggl[\frac{ r_{+}^{2} }{r_{+}^{2} + 2 \alpha} +  \alpha \left( \frac{D-5}{D-3} \right) \frac{1}{r_{+}^{2} + 2 \alpha} \Biggr] \\
\\
S &=& \frac{\mathcal{A}}{4 G} \left[ 1 + 2 \alpha
\left(\frac{D-2}{D-4}\right) \left( \frac{\mathcal{A}}{\Sigma_{D-2}} \right)^{-2/(D-2)} \right]
\end{eqnarray*}
 where $\mathcal{A} = \Sigma_{D-2} ~ r_+^{D-2}$ is the horizon area. The location of the horizon is found as roots of $q(r_+)=0$, where $q(r)=r^{D-3} + \alpha ~ r^{D-5} - \varpi$, and for the horizon to exist at all, one must also have $r_+^2 + 2 \alpha \geq 0$.

The highly damped quasi normal modes for the GB black holes (when $\omega_{I} \gg \omega_{R}$) has been worked out for $D=5$. These QNM frequencies are given by \cite{QnmGB}
\begin{eqnarray*}
\omega(n) \underset{n \rightarrow \infty}{\longrightarrow} T
\ln{Q} + i (2 \pi T) n
\end{eqnarray*}
(The imaginary part can be understood  in terms of a scattering matrix formalism; see e.g., \cite{tptirth}). We now use the Hod conjecture to obtain the entropy spacing for this spacetime. Accordingly, we identify as the relevant frequency $\omega_c$ the real part of $\omega$, i.e., we take $\omega_c = T \ln{Q}$. The entropy spacing is then given by
\begin{eqnarray*}
{S}_{n+1} - {S}_n = \frac{\partial {\mathcal
S}}{\partial M}   \omega_c= \ln Q
\end{eqnarray*}
Clearly the spacing $\Delta S\equiv{S}_{n+1} - {S}_n$ is a constant. This result depends
essentially only on the fact that ${\rm Re} ~ \omega_c \propto T$ leading to $(\partial S/\partial M)   \omega_c\ \propto T(\partial S/\partial M)$ which is a constant. For GB black holes the area is a function of entropy, ${\mathcal A}=F(S)$ which is \textit{not} linear. Hence,
for the area spectrum for this class of black holes we get
\begin{equation}
{\mathcal A}_{n+1} - {\mathcal A}_n = \frac{\partial {\mathcal
A}}{\partial M}   \omega_c =  g({\mathcal A}_n)\ln{Q}
\end{equation}
where, $g({\mathcal A}_n)=dF/dS$ is given by,
\begin{equation}
g({\mathcal A}_n) =  4\left[ 1 - 2 \alpha \left( \frac{\mathcal{A}_n}{\Sigma_3} \right)^{-2/3} \right]
\end{equation}
which is correct to ${\mathcal O}{(\alpha)}$. We therefore find that the entropy eigenvalues are discrete and equally spaced but the area spacing is \textit{not equidistant}. Hence, for GB gravity, the notion of \textit{quantum of entropy } is more natural than the {quantum of area}.

We shall now comment on the value of $\ln Q$ which determines the actual value of quantum of entropy.
Originally, in the case of Einstein gravity, the results of Bekenstein and Hod lead to the picture of a quantum black hole with the horizon built out of patches of area $\epsilon ~ A_P$, the most natural choice for $\epsilon$ being the (constant) spacing of eigenvalues of the area quantum operator, $\widehat{A}$. Hod further argued that one should take, for $\omega_c$, the real part of the quasi normal mode frequencies after the imaginary part has been sent to infinity. This leads to $\epsilon=4 \ln{k}$, $k$ being some integer. The numerical, as well as later analytical results for the quasi normal mode frequencies of spherically symmetric black holes in 3+1 dimensions, give $k=3$.

Recently, Maggiore \cite{mag} has put forth another argument which leads to identifying the transition frequency between large $n$ levels with the classical limit (rather than the real part of QNM frequencies, as was done by Hod). This gives $\epsilon=8 \pi$, consistent with earlier arguments of Bekenstein. While the specific value of area spacing is important for a statistical definition of entropy, it does not seem to be absolutely essential since, in a {\it semiclassical description}, the number of microstates need not exactly come out to be an integer, as was argued by Maggiore. Recently, all these arguments have been applied for the case of more general black holes in the context of Einstein gravity and it has been argued that in all such cases, when properly analyzed, one finds an equally spaced area spectrum \cite{medvid}.
The suggestion by Maggiore \cite{mag} to associate this classical limit with transition frequencies $(n+1) \rightarrow n$ as $n \rightarrow \infty$ leads to the replacement: $\ln{Q} \rightarrow 2 \pi$, which gives $\Delta S = 2 \pi$ (in units of $\hbar$). Thus, we obtain, for the quantum of entropy, a value of $2 \pi$ in agreement with the general arguments given earlier.

The broader picture which emerges from this analysis can be summarized along these lines: (a) In any theory which obeys the principle of equivalence, the gravitational field will be described at long wavelengths by a spacetime metric. (b) Around any event in spacetime, one can introduce a local inertial frame and --- by boosting with an acceleration --- a local Rindler frame. The observers using this coordinate system will have a local Rindler horizon with a temperature and entropy associated with the virtual deformations of the horizon. (c) \textit{Classically}, we interpret $A_{tot}$ in \eq{Sgrav} as the total entropy which is maximized for all the Rindler observers to give the field equations of the theory (which are the same as the equations of LL gravity). The on-shell value of the action giving the Wald entropy of the horizon which is interpreted as due to modes which are inaccessible to the given observer. (d) In the \textit{semiclassical} limit the $A_{tot}$ is interpreted as an action  and its value will affect the phase of the semiclassical wave function. (e) The observer independence of the semiclassical gravity requires this phase --- i.e., the Wald entropy of the horizon --- to be quantized in units of $2 \pi$. (Of course, mathematically, one could have treated $A_{tot}$ as the action functional even in classical theory.)

When we take the lowest order LL theory, we reproduce Einstein gravity and the quantization condition becomes equivalent to area quantization of the horizon as discussed several times in the literature. At the next order, we have the GB theory for which we have explicitly demonstrated the quantization of entropy. We believe that, once we have the structure of QNM in the case of LL theory --- which, as far as we know, has not yet been explicitly worked out --- the analysis given above can be repeated to give an explicit demonstration of this result. Since entropy is directly related with information content, the \textit{quantum of gravitational entropy} points out a new and intriguing relationship between gravity, quantum theory and thermodynamics.

DK and SS are supported by the Council of Scientific \& Industrial Research, India.


\begin{thebibliography}{99}

\bibitem{Bekenstein1}
J.D. Bekenstein, Lett. Nuovo Cimento 11, 467, (1974); J.D. Bekenstein [gr-qc/9710076].


\bibitem{chris}
Christodoulou, D, Phys. Rev. Lett. 25, 1596-1597, (1970).
\bibitem{comment}
This result only depends on spherical symmetry and the first law of black hole thermodynamics, and can be easily extended to black holes in Gauss-Bonnet gravity, leading to the same result for minimum increase of \textit{entropy}.


\bibitem{zeropoint}
B. S. DeWitt,\textit{ Phys. Rev. Lett.}, \textbf{13}, 114 (1964);
T. Padmanabhan \textit{ Ann. Phys.} (N.Y.), \textbf{165}, 38 (1985);
            \textit{Class. Quantum Grav.} \textbf{4}, L107 (1987);
A. Ashtekar et al., \textit{Phys. Rev. Lett.}, \textbf{69}, 237 (1992);
T. Padmanabhan  \textit{Phys. Rev. Lett.} \textbf{78},  1854 (1997) [hep-th/9608182];
              \textit{Phys. Rev.}  \textbf{D 57}, 6206 (1998);
L.J. Garay, \textit{ Int. J. Mod. Phys.} \textbf{A10}, 145 (1995).
\bibitem{lovelock}
C. Lanczos, Z.Phys. \textbf{73}, 147 (1932); Annals Math. \textbf{39}, 842 (1938);
D. Lovelock, Jour. Math. Phys., \textbf{12}, 498 (1971). For a recent
review, see Nathalie Deruelle and John Madore, [gr-qc/0305004], (2003).
\bibitem{gravtherm}
T Padmanabhan,{\it Class. Quan. Grav.}, {\bf 19}, 5387, (2002) [gr-qc/0204019];
T. Padmanabhan, {\it Gen. Rel. Grav.}, {\bf 34}, 2029, (2002) [gr-qc/0205090];
R-G Cai, L-M Cao, [gr-qc/0611071];
M Akbar, R-G Cai, [hep-th/0609128];
D. Kothawala, S. Sarkar, T. Padmanabhan,  Phys. Letts, \textbf{B 652}, 338-342 (2007) [gr-qc/0701002]
\bibitem{aseem-sudipta}
A Paranjape, S Sarkar and T Padmanabhan, {\it Phys. Rev.} {\bf D74},
                          104015, (2006) [hep-th/0607240].
\bibitem{ayan}
A Mukhopadhyay, T Padmanabhan, {\it Phys. Rev.} {\bf D74},124023, (2006).

\bibitem{noether}
 R. M. Wald, Phys. Rev. D \textbf{48}, 3427 (1993);
 V. Iyer, R. M. Wald, Phys. Rev. D \textbf{52}, 4430-4439 (1995).
\bibitem{paddypatel}
 T. Padmanabhan, Apoorva Patel [gr-qc/0309053], T. Padmanabhan,  Phys. Reports,  406, 49, (2005) [gr-qc/0311036]. The arguments were given for the case of Einstein's theory in these references but they continue to be valid for any gravitational action.

\bibitem{emergentpaddy}
T. Padmanabhan, AIP Conf. Proc. {\bf 939}, 114-123, (2007) [arXiv:0706.1654]
\bibitem{aseementropy}
 T. Padmanabhan,  \textit{Gen.Rel.Grav.},   \textbf{40}, 529-564 (2008) [arXiv:0705.2533];
 T. Padmanabhan, A. Paranjape, \textit{Phys.Rev.} \textbf{D75} 064004, (2007).

\bibitem{hod}
Shahar Hod, Phys. Rev. Lett. 81, 4293, (1998).
\bibitem{mag}
Maggiore, Michele, Phys. Rev. Lett. 100, 141301, (2008).
\bibitem{medvid}
Elias C. Vagenas [arXiv:0804.3264v2]; A.J.M. Medved [arXiv:0804.4346v2].
\bibitem{QnmGB}
Ramin G Daghigh et. al. .Class. Quantum Grav. {\bf 24}, 1981-1991, (2007).
Sayan K. Chakrabarti, Kumar S. Gupta. Int.J.Mod.Phys. {\bf A}, 21, 3565-3574 , (2006).
\bibitem{tptirth}
T. Padmanabhan, Class. Quan. Grav.,  \textbf{21}, L1 (2004) [gr-qc/0310027];
T.Roy Choudhury, T. Padmanabhan, Phys. Rev.D, \textbf{D 69} 064033 (2004) [gr-qc/0311064]


\bibitem{lovelockblack}
Jacobson T and Myers R C , Phys. Rev. Lett. {\bf 70}, 3684, (1993);
Myers R C and Simon J Z , Phys. Rev. D, {\bf 38}, 2434, (1998)




\end{thebibliography}
\end{document}